\documentclass[aps,twocolumn,preprintnumbers,nofootinbib,superscriptaddress]{revtex4-1}
\usepackage{times}
\usepackage{amsmath}
\usepackage{amssymb}
\usepackage{graphicx}
\usepackage{subfigure,accents}
\usepackage{booktabs}
\usepackage{rotating}
\usepackage{longtable}
\usepackage{bm}
\usepackage[pagebackref=false,colorlinks=,linkcolor=blue,citecolor=blue]{hyperref}
\usepackage{epstopdf}
 \pdfoutput=1
 \usepackage[utf8]{inputenc}
\usepackage{cases}
\usepackage{amsmath,float}
\usepackage{amssymb}
\usepackage{amsfonts}
\usepackage{amssymb}
\usepackage{dcolumn}
\usepackage{bm}
\usepackage{bbm}
\usepackage{graphicx}
\usepackage{xcolor}
\usepackage{array}
\usepackage{subfigure}
\usepackage{hyperref}
\usepackage{wasysym}

\newcommand{\be}{\begin{equation}}
\newcommand{\ee}{\end{equation}}
\newcommand{\ba}{\begin{eqnarray}}
\newcommand{\ea}{\end{eqnarray}}

\newcommand{\gsim}{\mathrel{\hbox{\rlap{\lower.55ex \hbox {$\sim$}}
                   \kern-.3em \raise.4ex \hbox{$>$}}}}
\newcommand{\lsim}{\mathrel{\hbox{\rlap{\lower.55ex \hbox {$\sim$}}
                   \kern-.3em \raise.4ex \hbox{$<$}}}}

\hypersetup{colorlinks=true,
            breaklinks=true,
            pdfstartview=Fit,
            linkcolor=blue,
            citecolor=blue,
            urlcolor=blue}
            
 \usepackage{graphicx}

\usepackage{amsmath}

\newcommand{\bw}{\begin{widetext}}
\newcommand{\ew}{\end{widetext}}

\def\b{\pmb{b}}

\newcommand{\inm}[1]{\ensuremath{\text{#1}}}
\newcommand{\diag}{\ensuremath{ \text{diag} }}
\newcommand{\lz}{\ensuremath{ {l_\inm{0}} }}

\newcommand{\EpMod}{\ensuremath{ \mathfrak{T} }}
\newcommand{\pr}{\ensuremath{ {p_\inm{r}} }}
\newcommand{\pt}{\ensuremath{ {p_\inm{t}} }}

\newcommand{\der}{\ensuremath{ d }}

\def\ber{\begin{eqnarray}}
\def\eer{\end{eqnarray}}
\def\beq{\begin{equation}}
\def\eeq{\end{equation}}

\newcommand{\dal}{\Box \phi}
\newcommand{\dpp}{\left(\nabla \phi \right)^2}

\renewcommand{\a}{\alpha}
\renewcommand{\b}{\beta}

\begin{document}

\title{Charged AdS black holes with finite electrodynamics in 4D Einstein-Gauss-Bonnet
gravity
}
\author{Kimet Jusufi}
\email{kimet.jusufi@unite.edu.mk}
\affiliation{Physics Department, State University of Tetovo, Ilinden Street nn, 
1200,
Tetovo, North Macedonia}

\begin{abstract}
    Using a modified expression for the electric potential in the context of T-duality (Gaete and Nicolini, Phys. Lett. B, 2022) we obtain an exact charged solution within 4D Einstein-Gauss-Bonnet (4D EGB) theory of gravity in the presence of a cosmological constant. We show that the solution exists in the regularized 4D EGB theory as well and we point out a correspondence between the black hole solution in 4D EGB  theory and the solution in the non-relativistic Horava–Lifshitz theory. The black hole solution is regular and free from singularity and as a special case we recover a class of well known solutions in the literature.
\end{abstract}
\maketitle

\section{Introduction}

The black hole singularity predicted by Einstein's theory is one of the main problems of general relativity. Different regular black hole solutions have been found in the literature. For example, the regular Bardeen black hole and the charged black hole obtained from T-duality  \cite{Nicolini:2019irw,Gaete:2022ukm} as well as black holes obtained in noncommutative geometry \cite{Nicolini}. Such solutions are described by regular geometry and the singularity predicted by classical general relativity is removed. In particular, one of the key concept used is the T-duality which basically identifies string theories on higher-dimensional spacetimes with mutually inverse compactification radii~$R$.  One has the replacement rules  
$R \rightarrow {R^\star}^2/R$ and $n \leftrightarrow w$, where $R^\star = \sqrt{\alpha'}$ is known as the self-dual radius, in addition $\alpha'$ represents the Regge slope, furthermore $n$ gives the Kaluza-Klein excitation and $w$ gives the winding number \cite{Nicolini}.  As noted by Padmanabhan \cite{Pad1} using the concept of T-duality one can find the propagator that
encodes string effects. This leads to a modified gravitational potential and electric potential \cite{Nicolini:2019irw,Gaete:2022une,Gaete:2022ukm}. On the other hand, the Einstein-Gauss-Bonnet theory is known to be topological in $4D$ since the Lagrangian in this theory is a total derivative and, as a result, it does not contribute to the gravitational dynamics. We therefore expect a non-trivial contribution  when $D \geq 5$. The problem of EGB theory in 4D was addressed by Glavan \& Lin \cite{Glavan:2019inb} with the main idea to rescale the Gauss-Bonnet coupling constant $\alpha$ as $\alpha/(D -4)$, and after taking the limit $D \to 4$ it was shown that a non-trivial gravitational dynamics is recovered. Interestingly, it was shown very soon that a regularized  4D theory at the level of action exists, namely as a special case of the 4D Horndeski theory \cite{Fernandes1,Hennigar:2020lsl}. For the static and spherically symmetric black hole solutions, it was shown that the regularized solution (obtained by rescaling the coupling constant) coincides with the original solution obtained by \cite{Glavan:2019inb}. In the present paper, we like to understand more about possible exact black hole exact solutions in 4D EGB theory inspired by the concept of T-duality.

This paper is organised as follows. In Sec. 2, we obtain a charged black hole solution in string T-Duality using the method of Glavan \& Lin \cite{Glavan:2019inb}. In Sec. 3, we obtain the same solution using a regularized method. In Sec. 4, we point our a connection with GUP principle. In Sec. 5, we study thermodynamics. Finally, in the Sec. 6, we comment on our result.

\section{Charged black hole solutions with finite electrodynamics in 4D EGB gravity}
Let us start by noting the momentum space propagator induced by the path integral duality (for the massless case) given by  \cite{Nicolini:2019irw}
\begin{equation}
G(k)= -\frac{\lz}{\sqrt{k^2}}\, K_1 \left(\lz \sqrt{k^2}\right),
\end{equation}
where $\lz$ denotes the zero-point length of spacetime and $K_1\!\left(x\right)$ is a modified Bessel function of the second kind. We have two special cases, first at small momenta, \textit{i.e.}~for ${(\lz k)}^2 \to 0$, we end up with the conventional massless propagator 
\begin{equation}
  G(k) = -k^{-2},\,\,\,\,\,\,\,\,\text{if}\,\,\,\,\,k^2 \ll 1/\lz^2.
\end{equation}
On the other hand, for large momenta we end up with 
\begin{equation}
G(k)
\sim -\lz^{1/2}\, {\left(k^2\right)}^{-3/4}\, e^{-\lz \sqrt{k^2}},\,\text{if}\,k^2 \gg 1/\lz^2.
\end{equation}
As we know, we can obtain the potential corresponding to the virtual-particle exchange via path integral of the force-mediating field. It was shown that for the potential of the system which consists of two pointlike masses, $m$ and $M$, at relative distance $\vec{r}$, one can write \cite{Nicolini:2019irw}
\begin{align}
\phi(r) \to -\frac{M}{\sqrt{r^2 + \lz^2}},
\label{eq:statpot}
\end{align}
In order to generate a new solution that includes the electric charge we need to take into account the energy contribution if the electric field encoded in the electromagnetic field given by \cite{Gaete:2022une,Gaete:2022ukm}
 \begin{eqnarray}
  V(r)=-\frac{Q}{\sqrt{r^2+l_0^2}}.
\end{eqnarray} 

Next, we can write the general action in EGB gravity in $D$-dimensions 
\begin{eqnarray}\label{action}
	S&=&\frac{1}{16 \pi}\int d^{D}x\sqrt{-g}\left[R+\frac{\alpha}{D-4} \mathcal{L}_{\text{GB}} \right]+S_{\text{matter}},
\end{eqnarray}
with $g$ being the determinant of the metric $g=\det g_{\mu\nu}$ and $\alpha$ is the Gauss-Bonnet coupling constant. We shall assume that for a physically acceptable solutions it has to be non-negative, i.e., $\alpha \ge 0$. Furthermore $\mathcal{L}_{\text{GB}}$ is the Lagrangian in GB gravity given by
\begin{equation}
	\mathcal{L}_{\text{GB}}=R^{\mu\nu\rho\sigma} R_{\mu\nu\rho\sigma}- 4 R^{\mu\nu}R_{\mu\nu}+ R^2\label{GB}.
\end{equation}
where $S_{\text{matter}}$ corresponds to the matter fields appearing in the theory. If we now vary the action (\ref{action}) with respect to metric $g_{\mu \nu}$ one can find \cite{Glavan:2019inb}
\begin{equation}\label{GBeq}
	G_{\mu\nu}+\frac{\alpha}{D-4} H_{\mu\nu}= 8 \pi \EpMod_{\mu \nu}.
\end{equation}
In the present work we would like to include the cosmological constant as well but we have included this term in the right hand side of the Einstein field equations as part of the energy-momentum tensor.  In total contribution is therefore given by the quantum corrected term, the electromagnetic field contribution, and the cosmological constant
\begin{eqnarray}
    \EpMod_{\mu \nu}=T^{str}_{\mu\nu}+T^{em}_{\mu\nu}+T^{vac}_{\mu\nu}
\end{eqnarray}
along with
\begin{eqnarray}
	G_{\mu\nu}&=&R_{\mu\nu}-\frac{1}{2}R g_{\mu\nu},\nonumber\\\notag
	H_{\mu\nu}&=&2\Bigr( R R_{\mu\nu}-2R_{\mu\sigma} {R}{^\sigma}_{\nu} -2 R_{\mu\sigma\nu\rho}{R}^{\sigma\rho} - R_{\mu\sigma\rho\delta}{R}^{\sigma\rho\delta}{_\nu}\Bigl)\\
	&-& \frac{1}{2}\mathcal{L}_{\text{GB}}g_{\mu\nu},\label{FieldEq}
\end{eqnarray}
where $R$ is the Ricci scalar, $R_{\mu\nu}$ is the Ricci tensor, $H_{\mu\nu}$ is known as the Lancoz tensor and, finally, we have the Riemann tensor  $R_{\mu\sigma\nu\rho}$.  Now the idea is to scale the coupling constant $ \alpha/(D-4)$, and by considering maximally symmetric spacetimes with curvature scale ${\cal K}$, we obtain \cite{Glavan:2019inb}
\begin{equation}\label{gbc}
\frac{g_{\mu\sigma}}{\sqrt{-g}} \frac{\delta \mathcal{L}_{\text{GB}}}{\delta g_{\nu\sigma}} = \frac{\alpha (D-2) (D-3)}{2(D-1)} {\cal K}^2 \delta_{\mu}^{\nu}.
\end{equation}
From the last equation we see that the variation of the GB action does not vanish in $D=4$ due to the re-scaled coupling constant which is the main idea in this method \cite{Glavan:2019inb}. The first contribution is due to the string quantum corrected tensor 
\begin{eqnarray}
   T^{str}_{\mu \nu}=-\frac{2}{\sqrt{-g}}\frac{\delta \left(\sqrt{-g} \mathcal{L}\right)}{\delta g^{\mu \nu}}.
\end{eqnarray}
The simplest choice is to take $\mathcal{L}=-\rho^{str}$, where the energy density of the modified matter using the Poisson equation \cite{Nicolini:2019irw}
\begin{equation}
\rho^{str}(r)= \frac{3 \lz^2 M}{4 \pi {\left(r^2 +\lz^2\right)}^{5/2}}.
\label{eq:rho}
\end{equation}

For the energy-momentum of the vacuum we have 
\begin{eqnarray}
    T^{vac}_{\mu\nu}=-\frac{\Lambda}{8\pi}\,\,\,g_{\mu \nu}
\end{eqnarray}
This has the form of a perfect fluid with 
\begin{equation}
    \rho^{vac}=-p^{vac}=\frac{\Lambda}{8 \pi}
\end{equation}
We will assume that the form of the energy-tensor the electromagnetic field has the form 
\begin{equation}
 T_{\mu \nu}^{em}=\frac{1}{4 \pi}\left(F_{\mu \sigma}{F_{\nu}}^{\sigma}    -\frac{1}{4}g_{\mu \nu}F_{\rho \sigma}F^{\rho \sigma}\right)
\end{equation}
where $F_{\mu\nu}=\nabla_\mu A_{\nu}-\nabla_\nu A_{\mu}$. Finally, from the total energy momentum we can write 
${(\EpMod^\mu}_\nu) = \diag\left(-\rho^{tot},\pr,\pt,\pt\right)$. Here, $\pr$ and $\pt$ are the radial and transverse pressures, respectively and total density $\rho^{tot}(r)=\rho^{str}(r)+\rho^{em}(r)+\rho^{vac}(r).$  The general solution in case of a static, spherically symmetric source reads
\begin{equation}
\label{eq:lineElem}
\der s^2
= -f(r) dt^2 +\frac{ dr^2}{f(r)} +r^2(d\theta^2+\sin^2\theta d\varphi^2).
\end{equation}
For the $r-r$ and $t-t$ components of the Einsteins field equations we obtain (see also for example \cite{Doneva:2020ped})
\begin{equation}
    -\frac{f'(r)}{r f(r)}+\frac{f^2(r)\alpha+(r^2-2 \alpha)f(r)-r^2+\alpha+8 \pi \rho r^4 }{r^2 f(r) (2 \alpha f(r)-r^2-2 \alpha)}=0
\end{equation}
\begin{equation}
    \frac{f'(r)}{r f(r)}+\frac{f^2(r)\alpha+(r^2-2 \alpha)f(r)-r^2+\alpha-8 \pi \rho r^4 }{r^2 f(r) (-2 \alpha f(r)+r^2+2 \alpha)}=0
\end{equation}

Considering the spherical symmetry existing in the spacetime metric imposes the only non-vanishing components of the Faraday tensor \cite{Gaete:2022ukm}
\begin{equation}
F_{tr}=-F_{rt}=-\frac{Q\, r}{\left(r^2+l_0^2\right)^{3/2}}.
\end{equation}
and the energy density of the energy-momentum field along with the pressure components are \cite{Gaete:2022ukm}
\begin{eqnarray}
    {\rho}^{em}(r)&=&-{p_r}^{em}(r)=\frac{Q^2 r^2}{8 \pi (r^2+l_0^2)^3}\\
    {p_t}^{em}(r)&=&\frac{Q^2 r^2}{8 \pi (r^2+l_0^2)^3}
\end{eqnarray}

One can then easily obtain the following exact solution
\begin{equation}
    f_{\pm}(r)=1+\frac{r^2}{2\alpha}\left(1 \pm \sqrt{1+\frac{8M\alpha}{(r^2+l_0^2)^{3/2}}-\frac{Q^2 \alpha \mathcal{F}(r) }{ (r^2+l_0^2)^2}-\frac{4 \alpha}{l^2}}\right),
\end{equation}
where 
\begin{eqnarray}
\mathcal{F}(r)&=&\frac{5}{2}+\frac{3 l_0^2}{2 r^2}-\frac{3 (r^2+l_0^2)^2 }{2 l_0 r^3}\arctan\left(\frac{r}{l_0}\right),
\end{eqnarray}
Considering a series expansion in $\alpha$ one can then easily obtain for the two branches of the solution
\begin{equation}
\begin{aligned}
    f_+(r)& = 1+\frac{2M r^2}{\left(r^2+l_0^2\right)^{3/2}}-\frac{5 Q^2 r^2}{8 (r^2+l_0^2)^2}\\
&-\frac{3 Q^2 l_0^2}{8 (r^2+l_0^2)^2}+\frac{3 Q^2}{8 l_0\,r }\arctan\left(\frac{r}{l_0}\right)-\frac{r^2}{l^2}+\frac{r^2}{\alpha},
\end{aligned}
\end{equation}
and
\begin{equation}
\begin{aligned}
    f_{-}(r)& = 1-\frac{2M r^2}{\left(r^2+l_0^2\right)^{3/2}}+\frac{5 Q^2 r^2}{8 (r^2+l_0^2)^2}\\
&+\frac{3 Q^2 l_0^2}{8 (r^2+l_0^2)^2}-\frac{3 Q^2}{8 l_0\,r }\arctan\left(\frac{r}{l_0}\right)+\frac{r^2}{l^2}.
\end{aligned}
\end{equation} 

As we shown in \cite{Gaete:2022ukm}, in the large limit $r>>l_0$, the ADM mass $\mathcal{M}$ is related to the mass parameter $M$ as follows
\begin{equation}\label{ADM}
    \mathcal{M}= M+\frac{3\pi Q^2}{32 l_0}.
\end{equation}  
we now rewrite the black hole solution as
\begin{equation}
    f(r)=1+\frac{r^2}{2\alpha}\left(1 \pm \sqrt{1+\frac{8  \mathcal{M} \alpha}{(r^2+l_0^2)^{3/2}}-\frac{Q^2 \alpha \mathcal{G}(r) }{ (r^2+l_0^2)^2}-\frac{4 \alpha}{l^2}}\right),
\end{equation}
where 
\begin{eqnarray}\notag
\mathcal{G}(r)&=&\frac{5}{2}+\frac{3 l_0^2}{2 r^2}+\frac{3\pi (r^2+l_0^2)^{1/2}}{16l_0}\\
&-&\frac{3 (r^2+l_0^2)^2 }{2 l_0 r^3}\arctan\left(\frac{r}{l_0}\right).
\end{eqnarray}
There are two branches of solutions as well and for physically acceptable solution one can choose the negative branch. In addition, we have the Maxwell's equation which is written as $\nabla_{\mu}F^{\mu \nu}=4 \pi J^{\nu}$, where $J^{\mu}=(\rho,\vec{j})$ is the four-current. One can check that for the black hole solution $\rho(r)= 3 \lz^2 Q/4 \pi {(r^2 +\lz^2)}^{5/2}$.  In what follows we are going to specialize our solution. \\

\begin{figure}
    \centering
    \includegraphics[width=8.6cm]{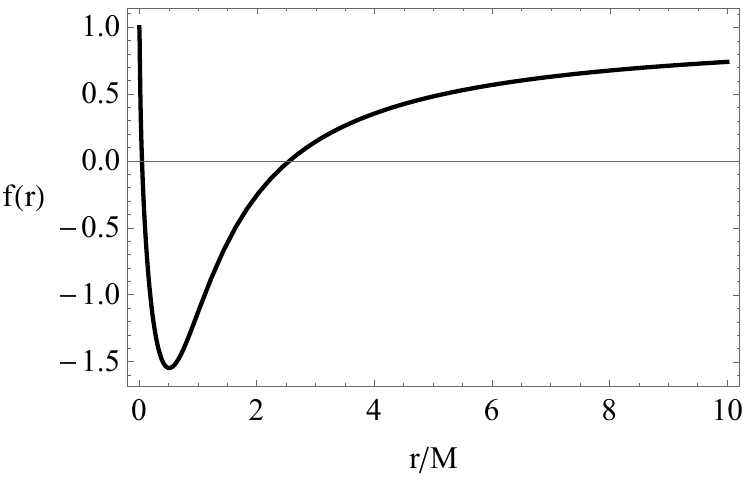}
    \caption{The metric function $f(r)$ as a function of $r$ for specific choice of parameters.}
    \label{fig:my_label}
\end{figure}

\subsection{Case: $\alpha \to 0$ (charged AdS black hole with string T-duality)}
In this limit as we saw one can take the negative branch of the solution and then rewrite it in terms of the ADM mass to obtain
\begin{equation}
    f(r)= 1-\frac{2\mathcal{M} r^2}{\left(r^2+l_0^2\right)^{3/2}}+\frac{Q^2 r^2}{(r^2+l_0^2)^2} \mathcal{H}(r)+\frac{r^2}{l^2},
\end{equation} 
where 
\begin{eqnarray}\notag
 \mathcal{H}(r)&=&\frac{5}{8}+\frac{3 l_0^2}{8 r^2}+\frac{3\pi (r^2+l_0^2)^{1/2}}{16 l_0}\\
 &-&\frac{3 (r^2+l_0^2)^2 }{8 l_0 r^3}\arctan\left(\frac{r}{l_0}\right).
\end{eqnarray}
In the limit of vanishing cosmological constant, one can check that this solution coincides with Gaete-Jusufi-Nicolini metric \cite{Gaete:2022ukm}. Interestingly, this solution can be viewed as a generalization of the well known Ayón-Beato--García solution obtained in the context of non-linear electrodynamics \cite{AB}. Furthermore in the limit $r>>l_0$, we obtain 
\begin{eqnarray}\notag
 \mathcal{H}(r)\sim \frac{5}{8}+\frac{3\pi r}{16 l_0}-\frac{3 r}{8 l_0}\arctan\left(\frac{r}{l_0}\right).
\end{eqnarray}

Introducing the relation $x=l_0/r<<1$ and using the identity $\arctan\left(1/x\right)+\arctan\left(x\right)=\pi/2$,
along with the fact that $\arctan(x)/x =1$ in the limit $x\to 0$. Now by taking the limit of the last equation the terms $3\pi r/16l_0$ cancels out and we get $\mathcal{H}\sim 1$. Finally, in this limit we obtain also the AdS RN solution
\begin{eqnarray}
f(r)=1-\frac{2\mathcal{M}}{r}+\frac{Q^2}{r^2}+\frac{r^2}{l^2}.\\
\end{eqnarray}

\subsection{Case: $Q \to 0$ (AdS Bardeen black hole)}
Yet another choice is to take the limit $Q\to 0$, in that case we obtain 
\begin{eqnarray}
    f(r)&=&1+\frac{r^2}{2\alpha}\left(1 \pm \sqrt{1+\frac{8M\alpha}{(r^2+l_0^2)^{3/2}}-\frac{4 \alpha}{l^2}}\right),
\end{eqnarray}
which has been reported in \cite{Singh:2020xju}. This solution was obtained again in the context of nonlinear electrodynamics.  

\subsection{Case: $l_0/r <<1$ (AdS charged black hole in 4D EGB  gravity ) }
Finally, let us consider the special case with $l_0/r <<1$. Introducing $x=l_0/r$ as in the above case and using the identities pointed out above, we obtain from metric (28) the following result
\begin{eqnarray}\notag
\mathcal{G}(r)\sim \frac{5}{2}+\frac{3\pi r}{16l_0}-\frac{3 r }{2 l_0 }\arctan\left(\frac{r}{l_0}\right).
\end{eqnarray}
Taking the limit of the last equation we get $\mathcal{G}\sim 4$, yielding in this way
\begin{equation}
    f(r)=1+\frac{r^2}{2\alpha}\left(1 \pm \sqrt{1+\frac{8\mathcal{M}\,\alpha}{r^3}-\frac{4 Q^2 \alpha }{r^4}-\frac{4 \alpha}{l^2}}\right).
\end{equation}
The last solution represents the charged AdS black hole in 4D EGB theory obtained in Ref. \cite{Fernandes:2020rpa}. \\

\section{Charged black hole solution in regularized 4D EGB gravity}
In the last section we saw that by taking the naive limit of the $D = 4$, the action (6) becomes ill-defined. It was shown that a well behaved action exists and it belongs to what is known as the scalar-tensor formulation of regularized 4D Einstein-Guass-Bonnet theory which can be written as  \cite{Hennigar:2020lsl,Fernandes1,Fernandes2}:
\begin{equation}
\begin{aligned}
S&=\frac{1}{16\pi} \int_{\mathcal{M}} d^{D} x \sqrt{-g}\Big[R+\alpha \big(4 G^{\mu \nu} \nabla_{\mu} \phi \nabla_{\nu} \phi-\phi \mathcal{G}\\
&+4 \square \phi(\nabla \phi)^{2}+2(\nabla \phi)^{4}\big) \Big]+S_{matter} \,.  \label{action}
\end{aligned}
\end{equation}%
 here $D=4$ and $S_{matter}$ encodes the effect of string corrections, finite electrodynamics, and cosmological constant being part of the energy-momentum tensor of the matter fields, respectively.  As was shown in \cite{Fernandes1,Hennigar:2020lsl} the regularization can be achieved by addition a counter-term of the form
\begin{eqnarray}
    -\frac{\alpha}{D-4}\int_{\mathcal{M}} d^{D}  \sqrt{-\tilde{g}} \tilde{\mathcal{G}}
\end{eqnarray}
where the tilde denotes quantities constructed from a conformal geometry $\tilde{g}_{\mu \nu} = g_{\mu \nu} e^{2\sigma}$.  Then by taking the limit $
 \lim_{D \to 4} (\int_{\mathcal{M}} d^{D} x \sqrt{-g} \mathcal{G}-\int_{\mathcal{M}} d^{D} x \sqrt{-\tilde{g}} \tilde{\mathcal{G}})/(D-4)$. If we now vary Eq.~(\ref{action}) with respect to metric we obtain  \cite{Hennigar:2020lsl,Fernandes1,Fernandes2}:
\begin{equation} \label{feqs0}
    G_{\mu \nu} + \alpha \mathcal{H}_{\mu \nu}=8\pi \, \EpMod_{\mu \nu}\, ,
\end{equation}
where one has \cite{Hennigar:2020lsl,Fernandes1,Fernandes2}
\begin{equation} 
\begin{aligned}
\mathcal{H}_{\mu\nu} &=2G_{\mu \nu} \dpp+4P_{\mu \alpha \nu \beta}\left(\nabla^\beta \nabla^\alpha \phi - \nabla^\alpha \phi \nabla^\beta \phi\right)\\
&+4\left(\nabla_\a \phi \nabla_\mu \phi - \nabla_\alpha \nabla_\mu \phi\right) \left(\nabla^\a \phi \nabla_\nu \phi - \nabla^\a \nabla_\nu \phi\right)\\
&+4\left(\nabla_\mu \phi \nabla_\nu \phi - \nabla_\nu \nabla_\mu \phi\right) \dal+g_{\mu \nu} \Big(2\left(\dal\right)^2 - \left( \nabla \phi\right)^4 \\
&+ 2\nabla_\b \nabla_\a\phi\left(2\nabla^\a \phi \nabla^\b \phi - \nabla^\b \nabla^\a \phi \right) \Big),
\end{aligned}
\end{equation}
and \cite{Fernandes2}
\begin{equation}
\begin{aligned}
P_{\alpha \beta \mu \nu}& = - R_{\alpha \beta \mu \nu}-g_{\alpha \nu}R_{\beta \mu}+g_{\alpha \mu}R_{\beta \nu}-g_{\beta \mu}R_{\alpha \nu}+g_{\beta \nu}R_{\alpha \mu}\\
&-\frac{1}{2}\left(g_{\alpha \mu}g_{\beta \nu}+g_{\alpha\nu}g_{\beta \mu}\right)R.
\end{aligned}
\end{equation}

On the other hand, if we vary the action with respect to $\phi$ it gives 
 \begin{equation} \label{scalareq}
\begin{aligned}
&R^{\mu \nu} \nabla_{\mu} \phi \nabla_{\nu} \phi - G^{\mu \nu}\nabla_\mu \nabla_\nu \phi - \dal \dpp +(\nabla_\mu \nabla_\nu \phi)^2\\
&- (\dal)^2 - 2\nabla_\mu \phi \nabla_\nu \phi \nabla^\mu \nabla^\nu \phi = \frac{1}{8}\mathcal{G}.
\end{aligned}
\end{equation}
Amazingly, one can use the above relations to find a simple closed form \cite{Fernandes1,Hennigar:2020lsl} 
\begin{equation}
    R+\frac{\alpha}{2}\mathcal{G} = -8\pi \,\EpMod.
    \label{Eq:trace}
\end{equation}

It was also shown that the action in the last equation is a special case of a more general well defined action nicely argued in \cite{Lu:2020iav}
\begin{eqnarray}\notag
S&=&\int d^D x\sqrt{-g}\Big[R
-2\Lambda+\alpha\Big(\phi\,\mathcal{G}+4G^{\mu\nu}\partial_\mu\phi\partial_\nu\phi \\\notag
&-&2\lambda Re^{-2\phi}-4(\partial\phi)^2\Box\phi+2((\partial\phi)^2)^2\\
&-&12\lambda(\partial\phi)^2e^{-2\phi}-6\lambda^2e^{-4\phi}\Big)\Big]+S_{matter}. 
\end{eqnarray}
In \cite{Lu:2020iav}, the authors used the Kaluza-Klein-like procedure to generate the limit of Gauss–Bonnet gravity by means of compactification (see also \cite{Kobayashi}). This action depends on a parameter $\lambda$ which describes the curvature of the (maximally symmetric) internal space \cite{Lu:2020iav}
\begin{equation}
R_{\mu \nu \sigma \beta}= \lambda(g_{\mu \sigma}g_{\nu \beta}-g_{\mu \beta}g_{\nu \sigma}),
\end{equation}
and the equivalence with is action (36) is achieved for vanishing $\lambda$. Assuming the black hole solution in terms of the metric 
\begin{equation}
\label{eq:lineElem}
\der s^2
= -f(r) dt^2 +\frac{ dr^2}{f(r)h(r)} +r^2(d\theta^2+\sin^2\theta d\varphi^2).
\end{equation}
We consider for simplicity the case of $\lambda=0$, then we can obtain the differential equations for the metric and the scalar field by inserting the above metric into the field equations (38) and (39). Alternatively, one can use the above metric ansatz to obtain the effective Lagrangian then by
varying w.r.t. $f, h, \phi$, we get the same equations of motion.
Note, however, we shall seek a solution for the metric function with $h=1$, when w.r.t. $f$ yields the equation \cite{Hennigar:2020lsl,Lu:2020iav}
\begin{eqnarray}
(\phi'^2+\phi'')(f r^2 \phi'^2-2 r f \phi'+f-1)=0.
\end{eqnarray}
The solution for the scalar field is \cite{Hennigar:2020lsl,Lu:2020iav,Fernandes2}
\begin{equation}
\phi'(r)=\pm \frac{1\pm \sqrt{f(r)}}{r\sqrt{f(r)}}.
\end{equation}
The remaining equation of motion from which we can find the metric for the static and spherically solution is (see also a similar approach \cite{Ovgun:2021ttv})
\begin{eqnarray}
\begin{aligned}
 &\frac{f'(r)}{r}-\frac{1-f(r)}{r^2}+ \alpha \left(\frac{2-2f(r)}{r^3} f'(r)  +\frac{(1-f(r))^2}{r^4}\right)\\
 &+\frac{Q^2\,r^2 }{\left(r^2+l_0^2\right)^{3}}+\frac{6 l_0^2 M}{\left(r^2+l_0^2\right)^{5/2}}-\frac{3}{l^2}=0.
 \end{aligned}
\end{eqnarray}
 Solving this differential equation and setting the constant of integration to zero, as expected, we get our solution given by
\begin{equation}
    f(r)=1+\frac{r^2}{2\alpha}\left(1 \pm \sqrt{1+\frac{8  M \alpha}{(r^2+l_0^2)^{3/2}}-\frac{Q^2 \alpha \mathcal{F}(r) }{ (r^2+l_0^2)^2}-\frac{4 \alpha}{l^2}}\right),
\end{equation}
where $\mathcal{F}(r)$ is given by Eq. (24).  If we now express this solution in terms of the ADM mass given by Eq. (27) we obtain final form of the solution given by Eq. (28). One should note here that going beyond the spherical symmetry, the naive limit of the $D = 4$ is not valid. 

Even though, the obtained solutions is in agreement with the regularized $4D$ EGB gravity, it was pointed out in \cite{Kobayashi,Aoki,Aoki:2020lig} the crucial problem in this theory is that the additional dynamical scalar field has to be infinitely strongly coupled if one tries to reproduce the results of \cite{Glavan:2019inb}. This mean the theory could not make any physical predictions in a reliable way and the results of \cite{Glavan:2019inb}, as well as many follow-up studies based on the regularized $4D$ EGB gravity are problematic because of the infinite strong coupling.  In order to reproduce the result of \cite{Glavan:2019inb}, authors in \cite{Aoki,Aoki:2020lig} pointed out that the only way to escape from this issue is by adding
appropriate counter term to the Hamiltonian \cite{Aoki,Aoki:2020lig} leading to the Lorentz violating theory. Specifically, this has been
made possible by breaking a part of diffeomorphism invariance which is consistent with the Lovelock theorem and, under reasonable assumptions, they argued how to construct a consistent theory of 4D EGB theory (see for details \cite{Aoki,Aoki:2020lig}).  In what follows, we will closely follow \cite{1} and argue that such an alternative regularization method of the 4D EGB via the 3+1 decomposition is closely linked to the solution in the Horava-Lifshitz theory which is a Lorentz violating theory. 

\subsection{Correspondence with the Horava-Lifshitz theory}
Using the formalism of the 3+1 decomposition, one can write the metric as \cite{1} 
\begin{equation}
ds^2=-N^2 dt^2+g_{ij}\left(dx^i+N^i dt\right)\left(dx^j+N^j
dt\right)\ ,
\end{equation}
along with the action in this theory \cite{1} 
\begin{eqnarray}\notag
 S&=&\int dt d^3 x
\sqrt{g}N\Big[\frac{2}{\kappa^2}\left(K_{ij}K^{ij}-\lambda
K^2\right)-\frac{\kappa^2}{2w^4}C_{ij}C^{ij}\\\notag
&+&\frac{\kappa^2
\mu}{2w^2}\epsilon^{ijk} R^{(3)}_{i\ell}
\nabla_{j}R^{(3)\ell}{}_k-\frac{\kappa^2\mu^2}{8} R^{(3)}_{ij} R^{(3)ij}\\
&+&\frac{\kappa^2
\mu^2}{8(1-3\lambda)} \frac{1-4\lambda}{4}(R^{(3)})^2+\mu^4 R^{(3)}\Big]+S_{matter},
\end{eqnarray}
where 
 \begin{equation}
 K_{ij}=\frac{1}{2N}\left(\dot{g}_{ij}-\nabla_i
N_j-\nabla_jN_i\right)\ ,
 \end{equation}
is the second fundamental form and
\begin{equation}
 C^{ij}=\epsilon^{ik\ell}\nabla_k
\left(R^{(3)j}{}_\ell-\frac{1}{4}R^{(3)} \delta^j_\ell\right)\ ,
 \end{equation}
is known as the Cotton tensor. If we consider a static and spherically symmetric background 
\begin{eqnarray}
ds^{2}=-N(r)^{2}dt^{2}+\frac{dr^{2}}{f(r)}+r^{2}(d\theta^{2}+\sin \theta^{2} d\phi^{2}),
\end{eqnarray}
by adding the matter fields in our case and closely following \cite{1}, for the Lagrangian it follows
\begin{eqnarray}\notag
{\cal{L}}&=&\frac{\kappa^2\mu^2}{8(1-3\lambda)}\frac{N}{\sqrt{f}}\Big((2\lambda-1)\frac{(f-1)^2}{r^2}
-2\lambda\frac{f-1}{r}f'\\
&+& \frac{\lambda-1}{2}f'^2-2 \omega
(1-f-rf')\Big)+\mathcal{L}_{matter},
\end{eqnarray}
where we defined $\omega=8 \mu^2(3\lambda-1)/\kappa^2$, along with 
\begin{equation}
    \mathcal{L}_{matter}=\mathcal{L}+\mathcal{L}_{e m}+\mathcal{L}_{\Lambda}.
\end{equation}
For the total Lagrangian we have to take into account the stringy corrections, electromagnetic field, and the cosmological constant, thus we get
\begin{equation}
\mathcal{L}_{matter}=\frac{N}{\sqrt{f}}\Big[-\gamma \frac{3 l_0^2 r^2 M}{4 \pi {\left(r^2 +\lz^2\right)}^{5/2}}-4\sigma \frac{Q^{2}r^4}{(r^{2}+l_0^2)^{3}}
+ 3 \delta \frac{r^2}{\ell^2}\Big],
\end{equation}
where $\gamma$, $\sigma$, $\delta$ and some constants that we need to fix. After we vary with respect to $N$ and we set $\lambda=1$, we get the following  equation of motion 
\begin{eqnarray}\notag
&&\frac{(f-1)^{2}}{r^{2}}-2 \frac{(f-1)}{r}f'- 2\omega(1-f-rf')\\
&+&2 \omega \frac{6 \lz^2 r^2 M}{{\left(r^2 +\lz^2\right)}^{5/2}}+2\omega \frac{Q^{2}r^4}{(r^{2}+l_0^2)^{3}}-2\omega \frac{3 r^2}{\ell^2}=0,
\end{eqnarray}
where we have used $\sigma=1/\kappa^2$, $\delta=4/\kappa^2$, $\gamma=4$ and $\mu^4 \kappa^2=2$. We now see that
by rescaling
\begin{eqnarray}
\omega\to \frac{1}{2\alpha},
\end{eqnarray}
we get 
\begin{eqnarray}\notag
&&\alpha\Big[\frac{(f-1)^{2}}{r^{2}}-2 \frac{(f-1)}{r}f'\Big]- (1-f-rf')\\
&+&\frac{6 \lz^2 r^2 M}{{\left(r^2 +\lz^2\right)}^{5/2}}+ \frac{Q^{2}r^4}{(r^{2}+l_0^2)^{3}}-\frac{3 r^2}{\ell^2}=0.
\end{eqnarray}
If we divide this equation with $r^2$, this equation is nothing but the differential equation (48), and therefore we get the same exact solution (49). This shows that there is a correspondence with the charged AdS solution in 4D EGB theory and Horava-Lifshitz theory. Finally, we point out that in the limit $l_0 \to 0$, we obtain the charged black hole where a similar correspondence was showin in \cite{Jusufi:2022ava}.

\begin{figure*}
    \centering
        \includegraphics[width=8.6cm]{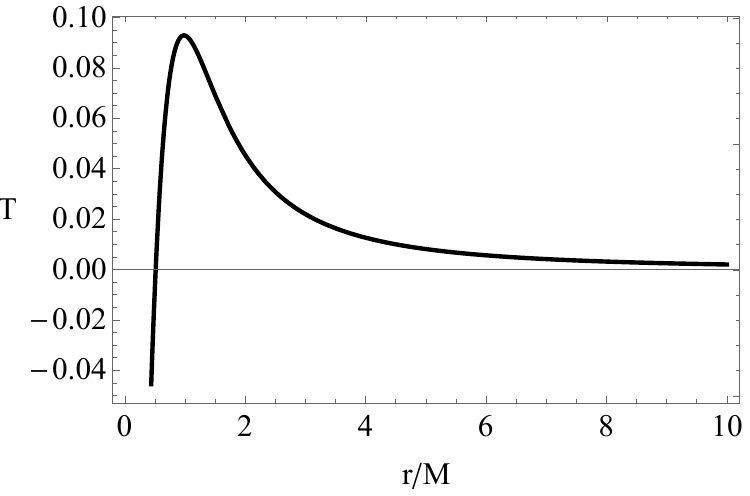}
          \includegraphics[width=8.6cm]{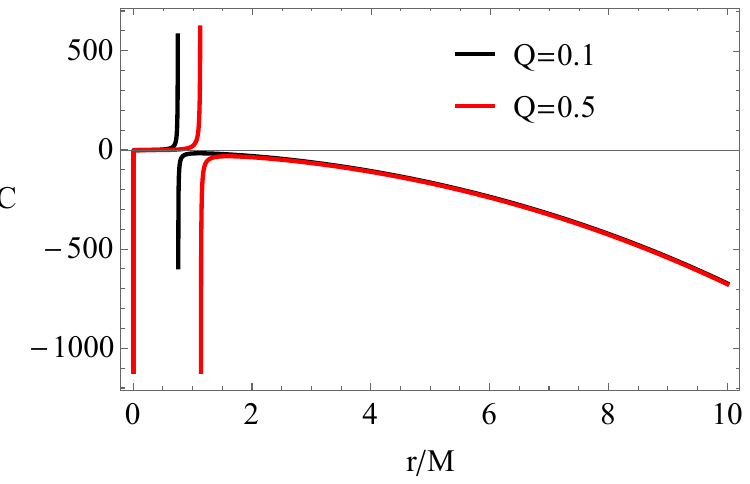}
    \caption{Left panel: The plot shows the Hawking temperature $T$ as a function of horizon $r=r_+$. Right panel: The plots for the the specific heat capacity using for specific choice of parameters.  }
    \label{fig:my_label}
\end{figure*}

\section{Connection with GUP}
In this section we shall show a close connection with the generalized uncertainty principle (GUP). Note that a connection between Bardeen black hole and GUP was shown in \cite{Maluf:2018ksj}. In the present work, we show a connection from another point of view based on the GUP principle for the black hole with modified ADM mass \cite{Carr1,Carr2}. This is another argument showing that the Bardeen regular solution can be viewed as a quantum corrected solution. In fact, the Bardeen solution is mostly considered as a solution of general relativity coupled to a nonlinear electrodynamics, but the string corrections as was shown lead to the same form of the metric \cite{Nicolini:2019irw}. Let us for simplicity assume a vanishing charge $Q=0$ and a vanishing cosmological constant $\Lambda=0$, respectively. Now let us recall that the GUP has been recently interpreted as the quantum mechanical hair to see this we can start with
\begin{eqnarray}
    \Delta x = \frac{1}{\Delta p}+\frac{\alpha_{GUP}^2\,l^2_{\text{Pl}}}{\hbar}\Delta p,
\end{eqnarray}
where $\alpha_{GUP}$ is the GUP parameter, $l_{\text{Pl}}=\sqrt{\hbar G/c^3} \sim 10^{-33}$ cm is the Planck length. Solving the $f(r_+)=0$ for mass we find
\begin{eqnarray}
       M=\frac{\left(r^2+l_0^2\right)^{3/2}}{2 r^2}|_{r_+},
\end{eqnarray}
and approximating to leading order terms we obtain
\begin{eqnarray}
    r_{\pm}\simeq M\pm \frac{\sqrt{4M^2-6 l_0^2}}{2}.
\end{eqnarray}

In the limit $l_0 \to 0$, we see that the ADM  mass gives $M_{ADM}=M$. This suggests that we can write 
\begin{eqnarray}
    r_+=2M_0,
\end{eqnarray}
where $M_0$ can be interpreted as the bare mass of the black hole. The ADM mass can then be written as 
\begin{eqnarray} \label{gupc}
    M_{ADM}=M_0\left(1+\frac{3\,l_0^2/4}{2 M_0^2}\right).
\end{eqnarray}
The second term on the right hand side can be interpreted as the quantum mechanical hair due to the GUP. We set $\Delta x \to R$, $\Delta p \to c M$, where $M_0$ is the mass forming an event horizon if it falls within its own Schwarzschild radius $R_S=2GM_0/c^2$ [here we have restored the constants $c$, $\hbar$ and $G$]. Using the $\beta$ (GUP corrected constant) formalism, it was shown in Ref. \cite{Carr1} that GUP has an important effect on the size of the black hole as follows
\begin{eqnarray}
    R_S'=\frac{2 G M_0}{c^2}\left(1+\frac{\beta}{2}\left(\frac{M_{\text{Pl}}}{M_0}\right)^2\right),\label{eq40}
\end{eqnarray}
where $M_{\text{Pl}}=\sqrt{\hbar c/G} \sim 10^{-5}$ g. With appropriate units and the scaling 
\begin{equation}
    \beta M^2_{\text{Pl}}\to \frac{3}{4}\,l_0^2,
\end{equation}
it can be easily seen that Eq. (\ref{eq40}) suggests a modified GUP ADM mass with $R_S'=2M_{ADM}$. For the ADM mass has been obtained \cite{Carr1,Carr2}
\begin{eqnarray}
       M_{ADM}=\frac{ G M_0}{c^2}\left(1+\frac{\beta}{2}\left(\frac{M_{\text{Pl}}}{M_0}\right)^2\right).
\end{eqnarray}

Recently, a similar connection was obtained by using the string solution obtained in \cite{Jusufi:2022uhk} by rescaling the string coupling parameter in a $D$-dimensional Callan-Myers-Perry black hole given by \cite{Jusufi:2022uhk} 
\begin{eqnarray}
    r_{\pm}= M\pm \frac{\sqrt{4M^2-2 \lambda}}{2}.
\end{eqnarray}
Hence we can see that $l_0=\sqrt{\lambda/3}$. This really shows the quantum origin of the Bardeen black hole. As was already pointed out in \cite{Gaete:2022ukm}, there exist a value of the mass such that for masses $M>M_{ext}$ there exists an outer event horizon but for $M<M_{ext}$, there spacetime has no horizon which can be viewed as a particle model.

 \section{Thermodynamics}
 The black hole solution given by Eq. (28) has an inner and outer event horizon as shown in Fig. 1. One can also compute the Hawking temperature of the black hole using
\begin{eqnarray}
    T=\frac{f'(r)}{4 \pi}|_{r_+}.
\end{eqnarray}
In Fig. 2 (left panel) we show the plot for the Hawking temperature for a specific domain of parameters.  We note that the negative values of temperature are not of physical interest. Equating $f(r_+)=0$ to zero, we get $M_{+}$ and $T_+$, respectively. Then by integrating the first law $dM=TdS$, the entropy for this black hole can be obtained as
\begin{eqnarray}
dS=\frac{1}{T}\bigg(\frac{\partial M_{+}}{\partial r_{+} }\bigg)_{Q,\alpha}dr_{+},
\end{eqnarray}
and the heat capacity can be evaluated via
\begin{equation}
C=\frac{\partial M}{\partial T}|_{r_+}=\frac{\partial M}{\partial r}\frac{\partial r}{\partial T}|_{r_+}.
\end{equation}
In general, we can evaluate these expressions numerically. Note here that if $C_+>0$ the black hole is said to be stable, and for $C_+ <0$, we say the black hole is unstable. Basically $C$ vanishes at some $r_=r_{ext}$ which means not all the values in the range given in Fig. 2 are allowed for $C$. Moreover there is singularity and a sign change which signals a phase transition at the maximal temperature.

\section{Conclusions}
 We have used the modified expression for the gravitational and electric potential inspired by the string T-duality we obtained an exact charged solution in the 4D EGB theory of gravity with a cosmological constant. It is shown that the solution exists in the regularized 4D EGB theory as well. We pointed out a problem regarding the additional dynamical scalar field which has to be infinitely strongly coupled if one tries to reproduce the results of 4d EGB theory. Following the arguments in \cite{Aoki,Aoki:2020lig} one can find a consistent theory and escape from this issue by adding appropriate counter term to the Hamiltonian \cite{Aoki,Aoki:2020lig} leading to the Lorentz violating theory by breaking a part of diffeomorphism invariance. We have used the 3+1 decomposition to show the correspondence between the solution in 4D EGB theory with the solution in the non-relativistic Horava–Lifshitz theory. The black hole solution reported here is regular and free from singularity and as a special case we obtain a class of well known solutions in the literature such as regular AdS Bardeen black hole, the AdS charged black hole in 4D EGB gravity, the AdS RN solution in general relativity and so on.  To this end, have investigated the thermodynamical aspects such as the Hawking radiation and the specific heat capacity.  We show a close connection between the quantum gravitational Bardeen black hole solution and the GUP principle.  In the near future, we shall study the phenomenological aspects of the solution reported in this work.

\acknowledgments{I would like to thank the referee for very helpful
and constructive comments which helped me to improve the paper
significantly.}

\end{document}